\documentstyle[12pt]{article}
\def\newpic#1{%
   \def\emline##1##2##3##4##5##6{%
      \put(##1,##2){\special{em:point #1##3}}%
      \put(##4,##5){\special{em:point #1##6}}%
      \special{em:line #1##3,#1##6}}}
\newpic{}

\begin{document}

\begin{center}
{\large
{\bf
ON THE TRIPLE CORRELATIONS \\IN HELICAL TURBULENCE}
\footnote{Slightly
extended version of the article "On the third moments in helical turbulence",
published in JETP Lett, 63(10), 802, 1996}
\\
\vskip 3mm

O.G.Chkhetiani}
\\
\vskip 3mm
{\it Space Research Institute, \\Russian Academy of Sciences,\\
Profsoyusnaya 84/32, Moscow, 117810, Russia\\
e--mail:ochkheti@mx.iki.rssi.ru
}
\end{center}


\begin{abstract}
The evolution of correlation characteristics in
homogeneous helical turbulence is considered. Additional
K\'{a}rm\'{a}n--Howarth type equations, describing the evolution of the
mixed correlation tensor of the velocity and vorticity are obtained. In the
helical scaling region, the solution of  obtained equation gives the exact
relation between antisymmetric component of a rank-three tensor and the
average dissipation of helicity. This relation is a helical analogue of
Kolmogorov's known ${\frac 45}$ law \cite{kolm41}.
\end{abstract}

Helicity plays an important role in the evolution and stability of turbulent
and laminar flows \cite{Moffat}. In \cite{Brissaud} the conception of the
helical cascade was introduced and the limiting cases of parallel energy and
helicity flows along the spectrum $E(k)\sim {\bar{\varepsilon}}%
^{2/3}k^{-5/3} $, $\ H(k)\sim {\bar{\eta}}{\bar{\varepsilon}}^{-1/3}k^{-5/3}$%
, corresponds to Kolmogorov' cascade, and a helicity flux with no energy
flux. $E(k)\sim {\bar{\eta}}^{2/3}k^{-7/3}$, $\ H(k)\sim {\bar{\eta}}%
^{2/3}k^{-4/3}$ --- pure helical cascade ($\bar{\varepsilon},\bar{\eta}$ -
dissipation of energy and helicity).

Many years pure helical cascade is considered as mainly theoretical
exercise without any practical significance. However, today we have nature
and laboratory results that is a direct evidence of minus seven-third law.
In turbulent MHD--flow in channel \cite{Branover}, tropical pretayphoon
atmosphere \cite{Sharkov} spectra of energy $E(k)\sim k^{-7/3}$,
quantitatively and qualitatively corresponds to the helical cascade
observe. In particular in tropical zone the helical cascade observes on
scales $l_h>6-7\> km$ \cite{Sharkov}).

Such a spectrum arises a problem of helicity generation. In \cite
{Krauze,Vainshtein} was shown, that for total helicity generation the
simultaneous presence of rigid--body rotation and temperature--dencity
stratification or other inhomogeneous factors be present simultaneously. In
\cite{Vainshtein} also was emphasized, that in this case helicity is pumping
in system on all turbulent scales. Helicity also generates in Eckman
boundary layers \cite{Hide}, in convective liquids \cite
{Lilly,Kurgansky,harfun} and at aerodynamic flow \cite{Hunt}. The
dissipative mechanism of helicity generation, first demonstrated in \cite
{Andre} should be noted also. Direct measurements have provided in turbulent
boundary layer, mixing layer \cite{Wallace90} and lattice turbulence \cite
{Kit87,Kit88,Wallace90,Dracos90,Kholmyansky91} are demonstrated the nonzero
values of mean helicity. Is noted also, that small helicity perturbations,
made by an artificial source before lattice are picked up and amplify
turbulence by a flow at lattice. Moreover, small, but the finite mean
helicity values at lattice were received even at the absence of an obvious
its source. Such phenomenon can be explained as influence of weak uncollinearity
of lattice apertures on the mirror symmetry of turbulent flow \cite{Wallace90}.

In \cite{harfun} helical cascades in a stratified and compressible turbulent
media were studied in detail. It was noted that there exists an inner
turbulent scale that separates the regions of Kolmogorov and helicity
cascades:
\[
l_h\sim {\frac{\bar{\varepsilon}}{\bar{\eta}}}.
\]
As rule, helicity cascade is observed on large scales (atmosphere) and for
more small scales we observe a Kolmogorov minus five--third law. However, if
the helicity source lies on small scales \cite{CGM}
we have reverse situation: helical scaling region lies on small scales ($%
k\leq l_h$) (see Figure 1).
\begin{figure}[tbp]
\unitlength=0.875mm
\special{em:linewidth 0.4pt}
\linethickness{0.4pt}
\begin{picture}(133.00,60.00)
\put(7.00,10.00){\vector(0,1){50.00}}
\put(7.00,10.00){\vector(1,0){50.00}}
\put(80.00,10.00){\vector(0,1){50.00}}
\put(80.00,10.00){\vector(1,0){50.00}}
\emline{32.00}{10.00}{1}{32.00}{12.00}{2}
\emline{32.00}{14.00}{3}{32.00}{16.00}{4}
\emline{32.00}{18.00}{5}{32.00}{20.00}{6}
\emline{32.00}{22.00}{7}{32.00}{24.00}{8}
\emline{32.00}{26.00}{9}{32.00}{28.00}{10}
\emline{32.00}{30.00}{11}{32.00}{32.00}{12}
\emline{32.00}{34.00}{13}{32.00}{36.00}{14}
\emline{105.00}{10.00}{15}{105.00}{12.00}{16}
\emline{105.00}{14.00}{17}{105.00}{16.00}{18}
\emline{105.00}{18.00}{19}{105.00}{20.00}{20}
\emline{105.00}{22.00}{21}{105.00}{24.00}{22}
\emline{105.00}{26.00}{23}{105.00}{28.00}{24}
\emline{105.00}{30.00}{25}{105.00}{32.00}{26}
\emline{105.00}{34.00}{27}{105.00}{36.00}{28}
\put(60.00,10.00){\makebox(0,0)[cc]{$k$}}
\put(133.00,10.00){\makebox(0,0)[cc]{$k$}}
\put(2.00,55.00){\makebox(0,0)[cc]{$E(k)$}}
\put(75.00,55.00){\makebox(0,0)[cc]{$E(k)$}}
\put(32.00,5.00){\makebox(0,0)[cc]{pretyphoon atmosphere}}
\put(105.00,5.00){\makebox(0,0)[cc]{laboratory MHD flow}}
\emline{11.00}{57.00}{29}{33.00}{35.00}{30}
\emline{32.00}{36.00}{31}{54.00}{23.00}{32}
\emline{105.00}{36.00}{33}{127.00}{14.00}{34}
\emline{83.00}{49.00}{35}{105.00}{36.00}{36}
\put(22.00,51.00){\makebox(0,0)[cc]{$-{7\over 3}$}}
\put(44.00,32.00){\makebox(0,0)[cc]{$-{5\over 3}$}}
\put(95.00,45.00){\makebox(0,0)[cc]{$-{5\over 3}$}}
\put(117.00,28.00){\makebox(0,0)[cc]{$-{7\over 3}$}}
\put(36.00,14.00){\makebox(0,0)[cc]{ $l_h^{-1}$}}
\put(109.00,14.00){\makebox(0,0)[cc]{$l_h^{-1}$}}
\end{picture}
\caption{}
\end{figure}
The scale invariance of characteristic functional allows to obtaine a form
of helical scaling in stratified turbulent fluid \cite{harfun}
\begin{eqnarray}
E(k) &=&\overline{\eta }^{2/3}k^{-7/3}\Psi \left( \left( kL_{*}\right)
^{1/2},kl_h\right) ,\quad  \\
E_{TT}(k) &=&N\overline{\eta }^{-1/3}k^{-4/3}\Psi _{TT}\left( \left(
kL_{*}\right) ^{1/2},kl_h\right)   \nonumber \\
L_{*} &=&l_h\left( l_h/L_{BO}\right) ^4,
\quad L_{BO}-\hbox{Bolgiano--Obukhov scale,}
\nonumber
\end{eqnarray}
and in compressible fluid \cite{harfun}
\begin{equation}
E(k) =\gamma ^{\frac{2\gamma }{(\gamma -1)(3\gamma -1)}}\rho
^{\frac{\gamma -1}{3\gamma -1}}C_0^{-\frac 2{3\gamma -1}}\overline{\eta }^{%
\frac{2\gamma }{3\gamma -1}}k^{-\frac{7\gamma -1}{3\gamma -1}}\Psi \left(
kl_h,kL\right),
\end{equation}
$$C_0-\hbox{ sound velocity.}$$
In the shock wave limit $\gamma \rightarrow 1$ we have $k^{-3}$
spectrum instead Kadomtsev--Petviashvili $k^{-2}$

\[
E(k)=e\overline{\eta }C_0^{-1}k^{-3}\Psi \left( kl_h,kL\right)
\]

The presence of helicity in turbulent system naturally puts a problem of
its influence to correlation properties of flow, in particular on high
correlations (and accordingly about possibility of its measurement). Not
doing of hypotheses about nature of the cascade we shall consider
evolution of the correlation characteristics in homogeneous helical
turbulence. At the initial moment of time $t=0$ exist nonzero average
helicity. Double correlation $\left\langle v_{1i}v_{2j}\right\rangle $ has a
form
\begin{equation}
\left\langle v_{1i}v_{2j}\right\rangle =A(r)\delta
_{ij}+B(r)r_ir_j+C\varepsilon _{ikl}r_l,\qquad r=|{\bf r}_2-{\bf r}_1|.
\end{equation}
Obviously
\[
\left\langle v^2\right\rangle =3A(0),\qquad \left\langle {\bf v}rot{\bf v}%
\right\rangle =-6C(0).
\]
We shall consider correlation of the third rank tensor
\[
b_{ik,l}=\left\langle v_{1i}v_{1k}v_{2l}\right\rangle =-\left\langle
v_{2i}v_{2k}v_{1l}\right\rangle
\]
General form of similar tensor following \cite{hince,chandras}
\[
b_{ik,l}=S_1(r)\delta _{ik}n_l+S_2(r)\delta _{il}n_k+S_3(r)\delta
_{nl}n_i+S_4(r)n_in_kn_l
\]
\begin{equation}
+S_5(r)\varepsilon _{ikl}+S_6(r)\varepsilon _{ikt}n_tn_l+S_7(r)\varepsilon
_{ilt}n_tn_k+S_8(r)\varepsilon _{klt}n_tn_i,
\end{equation}
Where ${\bf n}={\bf r}/|{\bf r}|$. The antisymmetric part tensor of the
third rank (proportional to $\varepsilon _{\ldots }$) usually was not
considered, as it differ from 0 only at helicity presence. Tensor $b_{ik,j}$
satisfies to the following conditions of symmetry \cite{hince,chandras}
\begin{equation}
b_{ik,j}=b_{ki,j},\quad b_{i,kj}=b_{i,jk},\quad b_{i,kj}=-b_{kj,i}
\end{equation}
And incompressibility condition
\begin{equation}
\frac{\partial b_{ik,j}}{\partial r_j}=0,
\end{equation}
Whence we receive, that
\[
S_5=S_6=0,\qquad S_7=S_8.
\]
Thus
\begin{eqnarray}
b_{ik,l} &=&D(r)\delta _{ik}n_l+E(r)(\delta _{il}n_k+\delta
_{nl}n_i)+F(r)n_in_kn_l  \nonumber \\
&&+S(r)\left( \varepsilon _{ilt}n_tn_k+\varepsilon _{klt}n_tn_i\right) ,
\end{eqnarray}
(The symmetric part of tensor is in detail considered in \cite
{landau,hince,chandras} and, as we shall see below, does not influence on
received results). We shall show, that the antisymmetric part, proportional $%
S(r)$, is directly connected with flow helicity. We shall consider mixed
point-to-point correlation tensor, component determined by product of
velocity and vorticity $\left\langle v_{1i}w_{2j}\right\rangle $, and having
in homogeneous isotropic case the following kind:
\begin{eqnarray}
\left\langle v_{1i}w_{2j}\right\rangle &=&-\left( 2C(r)+r\frac{\partial C(r)%
}{\partial r}\right) \delta _{ij}+{\frac 1r}\frac{\partial C(r)}{\partial r}%
r_ir_j  \nonumber \\
&&-\left( 5B(r)+r\frac{\partial B(r)}{\partial r}\right) \varepsilon
_{ijl}r_l.
\end{eqnarray}
It is easy to obtain the equations for its components
\begin{eqnarray}
{\frac \partial {\partial t}}\left\langle v_{1i}w_{2j}\right\rangle
&=&-\frac \partial {\partial x_{1l}}\left\langle
v_{1l}v_{1i}w_{2j}\right\rangle -\frac \partial {\partial
x_{1l}}\left\langle v_{1i}v_{2l}w_{2j}\right\rangle +\frac \partial
{\partial x_{1l}}\left\langle v_{1i}w_{2l}v_{2j}\right\rangle  \nonumber \\
&&-\frac \partial {\partial x_{1i}}\left\langle p_1w_{2j}\right\rangle +\nu
\Delta _1\left\langle v_{1i}w_{2j}\right\rangle +\nu \Delta _2\left\langle
v_{1i}w_{2j}\right\rangle
\end{eqnarray}
These equations, are similar to K\'{a}rm\'{a}n--Howarth equations and are
additional to them at the helicity presence in turbulent system. Owing to
solenoidality and regularity of correlation functions at $r=0$
\[
\left\langle p_1w_{2j}\right\rangle =0
\]
In the case of homogeneous turbulence we can extract derivative from
averaging braces
\begin{equation}
\frac \partial {\partial t}\left\langle h_{ij}\right\rangle =\varepsilon
_{jst}\frac{\partial ^2}{\partial r_l\partial r_s}\left[
b_{li,t}-b_{lt,i}\right] +2\nu \Delta _r\left\langle h_{ij}\right\rangle
\end{equation}
After simple transformations (the ''symmetric'' components tensor $b_{li,t}$
give the zero contribution) we receive
\begin{equation}
{\frac \partial {\partial t}}\left\langle h_{ij}\right\rangle =2\left\{
\left[ \frac{\partial ^2S}{\partial r^2}+{\frac 2r}{\frac{\partial S}{%
\partial r}}{\frac{6S}{r^2}}\right] n_in_j-\left[ {\frac{\partial ^2S}{%
\partial r^2}}+{\frac 4r}{\frac{\partial S}{\partial r}}\right] \delta
_{ij}+\nu \Delta \left\langle h_{ij}\right\rangle \right\} ,
\end{equation}
And for $i=j$ we receive
\begin{equation}
{\frac \partial {\partial t}}C(r)={\frac{2\nu }{r^4}}{\frac \partial
{\partial r}}\left( r^4{\frac \partial {\partial r}}C(r)\right) +{\frac
2{r^4}}{\frac \partial {\partial r}}\left( r^3S(r)\right) .
\end{equation}
Assuming, that at $r\to \infty $ the order of functions $\partial
C(r)/\partial r$ and $S(r)$ higher, than $1/r^4$ and $1/r^3$ accordingly,
and that $\nu r^4\partial C(r)/\partial r+r^3S(r)$ tends to zero at $r=0$,
we shall find a helical analogue of Loytsansky' invariant, which is
conserved on a final stage of free turbulence decay \cite{landau}
\begin{equation}
\Lambda _h=\int\nolimits_0^\infty C(r)r^4dr=\hbox{const}.
\end{equation}
This invariant concerns with the conservation of the mean product of the
vortex momentum and angular momentum $\displaystyle \approx \left\langle
{\bf P\cdot M}\right\rangle $.

Having presented $C(r)$ in form
\[
C(r)={\frac 16}\left( -\left\langle {\bf v}rot{\bf v}\right\rangle +\hat{C}%
(r)\right) ,
\]
We receive
\begin{equation}
{\bar{\eta}}+{\frac \partial {\partial t}}\hat{C}(r)={\frac{2\nu }{r^4}}{%
\frac \partial {\partial r}}\left( r^4{\frac \partial {\partial r}}\hat{C}%
(r)\right) +{\frac{12}{r^4}}{\frac \partial {\partial r}}\left(
r^3S(r)\right) ,
\end{equation}
Where $\displaystyle \overline{\eta }=2\nu \left\langle \frac{\partial v_i}{%
\partial x_j}\frac{\partial w_i}{\partial x_j}\right\rangle $ is the mean
helicity dissipation \cite{harfun}.

The helicity flow on spectrum is constant as in Kolmogorov', and in helical
interval \cite{Brissaud}. Therefore it is possible with sufficient accuracy
to neglect the change in time of $\hat{C}(r)$ in comparison with $\bar{\eta}$%
. Product on $r^4$ and integrate on $r$ we receive the following expression
for $S(r)$
\begin{equation}
S(r)={\frac{\bar{\eta}}{60}}r^2-\nu r{\frac \partial {\partial r}}\left(
C(r)\right) .
\end{equation}
It is obvious, that in depth of an inertial interval the viscous member is
small and we have simply
\begin{equation}
S(r)={\frac{\bar{\eta}}{60}}r^2.
\end{equation}
It's easy to obtaine the first terms of the Taylor expansion of $C(r)$ for
small $r$:
\begin{equation}
C(r)=-\frac 16\left( \left\langle {\bf v\cdot }curl\ {\bf v}\right\rangle +%
\frac{\overline{\eta }}{20\nu }r^2\right)
\end{equation}
This form of $C(r)$ turn to zero the expression for $S(r)$. Evidently, its
concerned with the more high order to $r$ dependence of $S(r)$ $(\sim r^4)$
in small--scale viscous region.

Received exact relation is similar to known '' $4/5"$ law, connecting
thriple longitudinal correlation of velocity with dissipation of energy \cite
{kolm41}. We shall note, that at its derivation we didn't use any
hypotheses about character of scaling. If in turbulent to system on which
or reasons arise helicity, occur additional nonzero components 2--pointed
correlation tensor of the third rank, proportional to the mean helicity
dissipation.

In particular for correlation $\left\langle v_i({\bf x})v_j({\bf x})w_j({\bf %
x+r})\right\rangle $ we receive
\begin{equation}
\left\langle v_i({\bf x})v_j({\bf x})w_j({\bf x+r})\right\rangle =-\frac{%
\overline{\eta }}{10}r_i.
\end{equation}
Having chosen axis $z$ parallel to a vector ${\bf r}$ we receive
\begin{eqnarray*}
b_{xy,z} &=&b_{y,xz}=0,\>\>b_{zy,x}=b_{yz,x}=-S(r), \\
b_{xz,y} &=&b_{z,xy}=S(r),\>\>b_{xy,z}=b_{y,xz}=0,\> \\
\>b_{zy,x} &=&b_{yz,x}=-S(r),\>\>b_{xz,y}=b_{zx,y}=S(r).
\end{eqnarray*}
The observation of such ''fine structure'' of high correlations of velocity
should be the direct and obvious evidence of helicity presence in turbulent
system.

It should be noted, that in \cite{harfun} expression for point-to-point
semiinvariant any rank in the helical scaling region is obtained
\begin{equation}
\left\langle v_{i1}({\bf x})v_{i2}({\bf x})\ldots v_{in}({\bf x}+{\bf r}%
)\right\rangle =\hbox{const}_n\cdot \left( \overline{\eta }r^2\right) ^{%
\frac{n}3} \Theta _{i_1,\ldots ,i_n}\left( {\bf n}\right),
\end{equation}
Where $\Theta_ {i_1,\ldots, i_n} \left ({\bf n} \right) $ - an angular part
of spectral tensor. (In nonhelical case accordingly $\left (\overline
{\varepsilon} r\right) ^ {\frac {n} 3} $).

In summary it would be necessary to express gratitude S.S.Moiseev and
A.V.Belyan for stimulating discussions. The work is carried out at support
RFFI (grant 96 - 02 - 19506), Russian Ministry of a Science, INTAS (grant
INTAS - 93 - 1194).

\end{document}